


\font\mayusc=cmcsc10 


      \font \ninebf                 = cmbx9
      \font \ninei                  = cmmi9
      \font \nineit                 = cmti9
      \font \ninerm                 = cmr9
      \font \ninesans               = cmss10 at 9pt
      \font \ninesl                 = cmsl9
      \font \ninesy                 = cmsy9
      \font \ninett                 = cmtt9
      \font \fivesans               = cmss10 at 5pt
						\font \sevensans              = cmss10 at 7pt
      \font \sixbf                  = cmbx6
      \font \sixi                   = cmmi6
      \font \sixrm                  = cmr6
						\font \sixsans                = cmss10 at 6pt
      \font \sixsy                  = cmsy6
      \font \tams                   = cmmib10
      \font \tamss                  = cmmib10 scaled 700
						\font \tensans                = cmss10
    
      \skewchar\ninei='177 \skewchar\sixi='177
      \skewchar\ninesy='60 \skewchar\sixsy='60
      \hyphenchar\ninett=-1
      \def\newline{\hfil\break}%
      \catcode`@=11
      \def\folio{\ifnum\pageno<\z@
      \uppercase\expandafter{\romannumeral-\pageno}%
      \else\number\pageno \fi}
      \catcode`@=12 

      \newfam\sansfam
      \textfont\sansfam=\tensans\scriptfont\sansfam=\sevensans
      \scriptscriptfont\sansfam=\fivesans
      \def\sans{\fam\sansfam\tensans}


      \def\petit{\def\rm{\fam0\ninerm}%
      \textfont0=\ninerm \scriptfont0=
\sixrm \scriptscriptfont0=\fiverm
       \textfont1=\ninei \scriptfont1=
\sixi \scriptscriptfont1=\fivei
       \textfont2=\ninesy \scriptfont2=
\sixsy \scriptscriptfont2=\fivesy
       \def\it{\fam\itfam\nineit}%
       \textfont\itfam=\nineit
       \def\sl{\fam\slfam\ninesl}%
       \textfont\slfam=\ninesl
       \def\bf{\fam\bffam\ninebf}%
       \textfont\bffam=\ninebf \scriptfont\bffam=\sixbf
       \scriptscriptfont\bffam=\fivebf
       \def\sans{\fam\sansfam\ninesans}%
       \textfont\sansfam=\ninesans \scriptfont\sansfam=\sixsans
       \scriptscriptfont\sansfam=\fivesans
       \def\tt{\fam\ttfam\ninett}%
       \textfont\ttfam=\ninett
       \normalbaselineskip=11pt
       \setbox\strutbox=\hbox{\vrule height7pt depth2pt width0pt}%
       \normalbaselines\rm


      \def\bvec##1{{\textfont1=\tbms\scriptfont1=\tbmss
      \textfont0=\ninebf\scriptfont0=\sixbf
      \mathchoice{\hbox{$\displaystyle##1$}}{\hbox{$\textstyle##1$}}
      {\hbox{$\scriptstyle##1$}}{\hbox{$\scriptscriptstyle##1$}}}}}


.

					\mathchardef\gammav="0100
     \mathchardef\deltav="0101
     \mathchardef\thetav="0102
     \mathchardef\lambdav="0103
     \mathchardef\xiv="0104
     \mathchardef\piv="0105
     \mathchardef\sigmav="0106
     \mathchardef\upsilonv="0107
     \mathchardef\phiv="0108
     \mathchardef\psiv="0109
     \mathchardef\omegav="010A


					\mathchardef\Gammav="0100
     \mathchardef\Deltav="0101
     \mathchardef\Thetav="0102
     \mathchardef\Lambdav="0103
     \mathchardef\Xiv="0104
     \mathchardef\Piv="0105
     \mathchardef\Sigmav="0106
     \mathchardef\Upsilonv="0107
     \mathchardef\Phiv="0108
     \mathchardef\Psiv="0109
     \mathchardef\Omegav="010A



\font\grbfivefm=cmbx5
\font\grbsevenfm=cmbx7
\font\grbtenfm=cmbx10 
\newfam\grbfam
\textfont\grbfam=\grbtenfm
\scriptfont\grbfam=\grbsevenfm
\scriptscriptfont\grbfam=\grbfivefm

\font\calbfivefm=cmbsy10 at 5pt
\font\calbsevenfm=cmbsy10 at 7pt
\font\calbtenfm=cmbsy10 
\newfam\calbfam
\textfont\calbfam=\calbtenfm
\scriptfont\calbfam=\calbsevenfm
\scriptscriptfont\calbfam=\calbfivefm



      \def\bvec#1{{\textfont1=\tams\scriptfont1=\tamss
      \textfont0=\tenbf\scriptfont0=\sevenbf
      \mathchoice{\hbox{$\displaystyle#1$}}{\hbox{$\textstyle#1$}}
      {\hbox{$\scriptstyle#1$}}{\hbox{$\scriptscriptstyle#1$}}}}



\def\pmbf#1{\leavevmode\setbox0=\hbox{#1}%
\kern-.02em\copy0\kern-\wd0
\kern.04em\copy0\kern-\wd0
\kern-.02em\copy0\kern-\wd0
\kern-.03em\copy0\kern-\wd0
\kern.06em\box0 }



						\def\monthname{%
   			\ifcase\month
      \or Jan\or Feb\or Mar\or Apr\or May\or Jun%
      \or Jul\or Aug\or Sep\or Oct\or Nov\or Dec%
   			\fi
							}%
					\def\timestring{\begingroup
   		\count0 = \time
   		\divide\count0 by 60
   		\count2 = \count0   
   		\count4 = \time
   		\multiply\count0 by 60
   		\advance\count4 by -\count0   
   		\ifnum\count4<10
     \toks1 = {0}%
   		\else
     \toks1 = {}%
   		\fi
   		\ifnum\count2<12
      \toks0 = {a.m.}%
   		\else
      \toks0 = {p.m.}%
      \advance\count2 by -12
   		\fi
   		\ifnum\count2=0
      \count2 = 12
   		\fi
   		\number\count2:\the\toks1 \number\count4 \thinspace \the\toks0
					\endgroup}%

				\newskip\abovelistskip      \abovelistskip = .5\baselineskip 
				\newskip\interitemskip      \interitemskip = 0pt
				\newskip\belowlistskip      \belowlistskip = .5\baselineskip
				\newdimen\listleftindent    \listleftindent = 0pt
				\newdimen\listrightindent   \listrightindent = 0pt

				%


\def\petit{\def\rm{\fam0\ninerm}%
\textfont0=\ninerm \scriptfont0=\sixrm \scriptscriptfont0=\fiverm
\textfont1=\ninei \scriptfont1=\sixi \scriptscriptfont1=\fivei
\textfont2=\ninesy \scriptfont2=\sixsy \scriptscriptfont2=\fivesy
       \def\it{\fam\itfam\nineit}%
       \textfont\itfam=\nineit
       \def\sl{\fam\slfam\ninesl}%
       \textfont\slfam=\ninesl
       \def\bf{\fam\bffam\ninebf}%
       \textfont\bffam=\ninebf \scriptfont\bffam=\sixbf
       \scriptscriptfont\bffam=\fivebf
       \def\sans{\fam\sansfam\ninesans}%
       \textfont\sansfam=\ninesans \scriptfont\sansfam=\sixsans
       \scriptscriptfont\sansfam=\fivesans
       \def\tt{\fam\ttfam\ninett}%
       \textfont\ttfam=\ninett
       \normalbaselineskip=11pt
       \setbox\strutbox=\hbox{\vrule height7pt depth2pt width0pt}%
       \normalbaselines\rm
      \def\vec##1{{\textfont1=\tbms\scriptfont1=\tbmss
      \textfont0=\ninebf\scriptfont0=\sixbf
      \mathchoice{\hbox{$\displaystyle##1$}}{\hbox{$\textstyle##1$}}
      {\hbox{$\scriptstyle##1$}}{\hbox{$\scriptscriptstyle##1$}}}}}

      \def\footnoterule{\kern-3pt\hrule width 2cm\kern2.6pt}
      \newdimen\oldparindent\oldparindent=1.5em
      \parindent=1.5em
 
\newcount\footcount \footcount=0
\def\advftncnt{\advance\footcount by1\global\footcount=\footcount}
      \def\fnote#1{\advftncnt$^{\the\footcount}$\begingroup\petit
      \parfillskip=0pt plus 1fil
      \def\textindent##1{\hangindent0.5\oldparindent\noindent\hbox
      to0.5\oldparindent{##1\hss}\ignorespaces}%
 \vfootnote{$^{\the\footcount}$}
{#1\nullbox{0mm}{2mm}{0mm}\vskip-9.69pt}\endgroup}


      \def\item#1{\par\noindent
      \hangindent6.5 mm\hangafter=0
      \llap{#1\enspace}\ignorespaces}
      
      \def\leaderfill{\kern0.5em\leaders
\hbox to 0.5em{\hss.\hss}\hfill\kern
      0.5em}
						\def\hb{\hfill\break}

    \def\centerrule#1{\centerline{\kern#1\hrulefill\kern#1}}


      \def\boxit#1{\vbox{\hrule\hbox{\vrule\kern3pt
						\vbox{\kern3pt#1\kern3pt}\kern3pt\vrule}\hrule}}

      \def\tightboxit#1{\vbox{\hrule\hbox{\vrule
						\vbox{#1}\vrule}\hrule}}

      \def\looseboxit#1{\vbox{\hrule\hbox{\vrule\kern5pt
						\vbox{\kern5pt#1\kern5pt}\kern5pt\vrule}\hrule}}

      \def\youboxit#1#2{\vbox{\hrule\hbox{\vrule\kern#2
						\vbox{\kern#2#1\kern#2}\kern#2\vrule}\hrule}}



			\def\whitetile#1#2#3{\setbox0=\null
			\ht0=#1 \dp0=#2\wd0=#3 \setbox1=
\hbox{\tightboxit{\box0}}\lower#2\box1}

			\def\nullbox#1#2#3{\setbox0=\null
			\ht0=#1 \dp0=#2\wd0=#3\box0}




\def\fig{\leavevmode Fig.}

\def\equ{\leavevmode Eq.}

\def\sect{\leavevmode Sect.}

\def\equn#1{\ifmmode \eqno{\rm #1}\else \equ~#1\fi}



\def\tev{\ifmmode \mathop{\rm TeV}\nolimits\else {\rm TeV}\fi}
\def\gev{\ifmmode \mathop{\rm GeV}\nolimits\else {\rm GeV}\fi}
\def\mev{\ifmmode \mathop{\rm MeV}\nolimits\else {\rm MeV}\fi}
\def\kev{\ifmmode \mathop{\rm keV}\nolimits\else {\rm keV}\fi}
\def\ev{\ifmmode \mathop{\rm eV}\nolimits\else {\rm eV}\fi}

\def\chidof{\ifmmode
\mathop\chi^2/{\rm d.o.f.}\else $\chi^2/{\rm d.o.f.}\null$\fi}

\def\msbar{\ifmmode
\mathop{\overline{\rm MS}}\else$\overline{\rm MS}$\null\fi}


\def\physmatex{P\kern-.14em\lower.5ex\hbox{\sevenrm H}ys
\kern -.35em \raise .6ex \hbox{{\sevenrm M}a}\kern -.15em
 T\kern-.1667em\lower.5ex\hbox{E}\kern-.125emX\null}%

\def\ref#1{$^{[#1]}$\relax}

\def\ajnyp#1#2#3#4#5{
\frenchspacing{\mayusc #1}, {\sl#2}, {\bf #3} ({#4}) {#5}}













\def\ddal{\mathop{\vrule height 7pt depth0.2pt
\hbox{\vrule height 0.5pt depth0.2pt width 6.2pt}
\vrule height 7pt depth0.2pt width0.8pt
\kern-7.4pt\hbox{\vrule height 7pt depth-6.7pt width 7.pt}}}
\def\sdal{\mathop{\kern0.1pt\vrule height 4.9pt depth0.15pt
\hbox{\vrule height 0.3pt depth0.15pt width 4.6pt}
\vrule height 4.9pt depth0.15pt width0.7pt
\kern-5.7pt\hbox{\vrule height 4.9pt depth-4.7pt width 5.3pt}}}
\def\ssdal{\mathop{\kern0.1pt\vrule height 3.8pt depth0.1pt width0.2pt
\hbox{\vrule height 0.3pt depth0.1pt width 3.6pt}
\vrule height 3.8pt depth0.1pt width0.5pt
\kern-4.4pt\hbox{\vrule height 4pt depth-3.9pt width 4.2pt}}}




\mathchardef\lap='0001


\def\lsim{\mathop{\setbox0=\hbox{$\displaystyle 
\raise2.2pt\hbox{$\;<$}\kern-7.7pt\lower2.6pt\hbox{$\sim$}\;$}
\box0}}
\def\gsim{\mathop{\setbox0=\hbox{$\displaystyle 
\raise2.2pt\hbox{$\;>$}\kern-7.7pt\lower2.6pt\hbox{$\sim$}\;$}
\box0}}

\def\gsimsub#1{\mathord{\vtop to0pt{\ialign{##\crcr
$\hfil{{\mathop{\setbox0=\hbox{$\displaystyle 
\raise2.2pt\hbox{$\;>$}\kern-7.7pt\lower2.6pt\hbox{$\sim$}\;$}
\box0}}}\hfil$\crcr\noalign{\kern1.5pt\nointerlineskip}
$\hfil\scriptstyle{#1}{}\kern1.5pt\hfil$\crcr}\vss}}}

\def\lsimsub#1{\mathord{\vtop to0pt{\ialign{##\crcr
$\hfil\displaystyle{\mathop{\setbox0=\hbox{$\displaystyle 
\raise2.2pt\hbox{$\;<$}\kern-7.7pt\lower2.6pt\hbox{$\sim$}\;$}
\box0}}
\def\gsim{\mathop{\setbox0=\hbox{$\displaystyle 
\raise2.2pt\hbox{$\;>$}\kern-7.7pt\lower2.6pt\hbox{$\sim$}\;$}
\box0}}\hfil$\crcr\noalign{\kern1.5pt\nointerlineskip}
$\hfil\scriptstyle{#1}{}\kern1.5pt\hfil$\crcr}\vss}}}

\def\dd{{\rm d}}





\def\frac#1#2{{#1\over#2}}
\def\dfrac#1#2{{\displaystyle{#1\over#2}}}
\def\tfrac#1#2{{\textstyle{#1\over#2}}}
\def\ffrac#1#2{\leavevmode
   \kern.1em \raise .5ex \hbox{\the\scriptfont0 #1}%
   \kern-.1em $/$%
   \kern-.15em \lower .25ex \hbox{\the\scriptfont0 #2}%
}%



\def\brochureb#1#2#3{\pageno#3
\headline={\ifodd\pageno{\rheadline}
\else\lheadline\fi}
\def\rheadline{\hfil -{#2}-\hfil}
\def\lheadline{\hfil-{#1}-\hfil}
\footline={\hss -- \number\pageno\ --\hss}
\voffset=2\baselineskip}

\def\nada{\phantom{M}\kern-1em}
\def\brochureendcover#1{\vfill\eject\pageno=1{\nada#1}\vfill\eject}





\def\chapterb#1#2#3{\pageno#3
\headline={\ifodd\pageno{\ifnum\pageno=#3\hfil\else\rheadline\fi}
\else\lheadline\fi}
\def\rheadline{\hfil -{#2}-\hfil}
\def\lheadline{\hfil-{#1}-\hfil}
\footline={\hss -- \number\pageno\ --\hss}
\voffset=2\baselineskip}


\def\bookendchapter{\ifodd\pageno
 \vfill\eject\footline={\hfill}\headline={\hfill}\null \vfill\eject
 \else\vfill\eject \fi}

\def\obookendchapter{\ifodd\pageno\vfill\eject
 \else\vfill\eject\null \vfill\eject\fi}


\def\booksection#1{
\setbox0=\vbox{\hsize=0.85\hsize\tolerance=500\raggedright\hfuzz=6mm
\noindent{\medfib #1}\medskip}\goodbreak\vskip0.6cm\box0
\nobreak
\noindent}
\def\booksubsection#1{
\setbox0=\vbox{\hsize=0.85\hsize\tolerance=400\raggedright\hfuzz=4mm
\noindent{\fib #1}\smallskip}\goodbreak\vskip0.45cm\box0
\nobreak
\noindent}




\def\figurasc#1#2{\petit{\noindent\sc#1}\ #2}

\def\captiontype{\tolerance=800\hfuzz=1mm\raggedright\noindent}



\def\abstracttype#1{
\hsize0.7\hsize\tolerance=800\hfuzz=0.5mm \noindent{\fib #1}\par
\medskip\petit}


\def\hb{\hfill\break}


\font\twelverm=cmr12 
\font\smallsc=cmcsc10 at 9pt 
\font\fib=cmfib8
\font\medfib=cmfib8 at 9pt


\font\sc=cmcsc10 

\font\addressfont=cmbxti10 at 9pt


\catcode`@=11 

\newdimen\pagewidth \newdimen\pageheight \newdimen\ruleht
 \maxdepth=2.2pt  \parindent=3pc
\pagewidth=\hsize \pageheight=\vsize \ruleht=.4pt
\abovedisplayskip=6pt plus 3pt minus 1pt
\belowdisplayskip=6pt plus 3pt minus 1pt
\abovedisplayshortskip=0pt plus 3pt
\belowdisplayshortskip=4pt plus 3pt

\newinsert\margin
\dimen\margin=\maxdimen




\newdimen\paperheight \paperheight = \vsize
\def\topmargin{\afterassignment\@finishtopmargin \dimen0}%
\def\@finishtopmargin{%
  \dimen2 = \voffset		
  \voffset = \dimen0 \advance\voffset by -1in
  \advance\dimen2 by -\voffset	
  \advance\vsize by \dimen2	
}%
\def\advancetopmargin{%
  \dimen0 = 0pt \afterassignment\@finishadvancetopmargin \advance\dimen0
}%
\def\@finishadvancetopmargin{%
  \advance\voffset by \dimen0
  \advance\vsize by -\dimen0
}%
\def\bottommargin{\afterassignment\@finishbottommargin \dimen0}%
\def\@finishbottommargin{%
  \@computebottommargin		
  \advance\dimen2 by -\dimen0	
  \advance\vsize by \dimen2	
}%
\def\advancebottommargin{%
  \dimen0 = 0pt\afterassignment\@finishadvancebottommargin \advance\dimen0
}%
\def\@finishadvancebottommargin{%
  \advance\vsize by -\dimen0
}%
\def\@computebottommargin{%
  \dimen2 = \paperheight	
  \advance\dimen2 by -\vsize	
  \advance\dimen2 by -\voffset	
  \advance\dimen2 by -1in	
}%
\newdimen\paperwidth \paperwidth = \hsize
\def\leftmargin{\afterassignment\@finishleftmargin \dimen0}%
\def\@finishleftmargin{%
  \dimen2 = \hoffset		
  \hoffset = \dimen0 \advance\hoffset by -1in
  \advance\dimen2 by -\hoffset	
  \advance\hsize by \dimen2	
}%
\def\advanceleftmargin{%
  \dimen0 = 0pt \afterassignment\@finishadvanceleftmargin \advance\dimen0
}%
\def\@finishadvanceleftmargin{%
  \advance\hoffset by \dimen0
  \advance\hsize by -\dimen0
}%
\def\rightmargin{\afterassignment\@finishrightmargin \dimen0}%
\def\@finishrightmargin{%
  \@computerightmargin		
  \advance\dimen2 by -\dimen0	
  \advance\hsize by \dimen2	
}%
\def\advancerightmargin{%
  \dimen0 = 0pt \afterassignment\@finishadvancerightmargin \advance\dimen0
}%
\def\@finishadvancerightmargin{%
  \advance\hsize by -\dimen0
}%
\def\@computerightmargin{%
  \dimen2 = \paperwidth		
  \advance\dimen2 by -\hsize	
  \advance\dimen2 by -\hoffset	
  \advance\dimen2 by -1in	
}%

\def\onepageout#1{\shipout\vbox{ 
    \offinterlineskip 
    \vbox to 3pc{ 
      \iftitle 
        \global\titlefalse 
        \setcornerrules 
      \else\ifodd\pageno \rightheadline\else\leftheadline\fi\fi
      \vfill} 
    \vbox to \pageheight{
      \ifvoid\margin\else 
        \rlap{\kern31pc\vbox to\z@{\kern4pt\box\margin \vss}}\fi
      #1 
      \ifvoid\footins\else 
        \vskip\skip\footins \kern-3pt
        \hrule height\ruleht width\pagewidth \kern-\ruleht \kern3pt
        \unvbox\footins\fi
      \boxmaxdepth=\maxdepth
      } 
    }
  \advancepageno}

\def\setcornerrules{\hbox to \pagewidth{\vrule width 1pc height\ruleht
    \hfil \vrule width 1pc}
  \hbox to \pagewidth{\llap{\sevenrm(page \folio)\kern1pc}%
    \vrule height1pc width\ruleht depth\z@
    \hfil \vrule width\ruleht depth\z@}}
\newbox\partialpage






\input epsf.sty
\raggedbottom
\footline={\hfill}
\rightline{FTUAM 01-01}
\rightline{UG-FT-126/00}
\rightline{hep-ph/0102247}
\rightline{January, 2001}
\bigskip
\hrule height .3mm
\vskip.6cm
\centerline{{\twelverm Improved Calculation of $F_2$ in 
Electroproduction and $xF_3$ in Neutrino Scattering to NNLO}}
\medskip
\centerline{{\twelverm and 
 Determination of  $\alpha_s$}}
\medskip
\centerrule{.7cm}
\vskip1cm
\setbox8=\vbox{\hsize65mm {\noindent\fib J. Santiago} 
\vskip .1cm
\noindent{\addressfont Departamento de F\'\i sica Te\'orica\hb y del Cosmos,\hb
Universidad de Granada,\hb
E-18071, Granada, Spain.}}
\centerline{\box8}
\medskip
\setbox7=\vbox{\hsize65mm \fib and} 
\centerline{\box7}
\medskip
\setbox9=\vbox{\hsize65mm {\noindent\fib F. J. 
Yndur\'ain} 
\vskip .1cm
\noindent{\addressfont Departamento de F\'{\i}sica Te\'orica, C-XI,\hb
 Universidad Aut\'onoma de Madrid,\hb
 Canto Blanco,\hb
E-28049, Madrid, Spain.}\hb}
\smallskip
\centerline{\box9}
\bigskip
\setbox0=\vbox{\abstracttype{Abstract}We improve the existing calculations of 
deep inelastic scattering to next to next to leading order in 
the following manner. First, we use the recently calculated values of
 the anomalous dimensions for moments with index 
$n=10,12$ in $ep$ scattering. Second, we use 
also recently calculated anomalous diemensions for $n=1,3,\dots,13$ 
for $xF_3$ in $\nu N$ scattering, to extend the calculation to 
this process. 
Lastly, we 
use the determinantal constraints on the moments that follow 
from positivity of the structure functios to stabilize the fits. 
We find substantial improvement over previous results, getting the following numbers:
$$\eqalign{\lambdav(n_f=4, 3\; {\rm loop})=&274\pm20\;\mev;
\quad\alpha_s^{3\;{\rm loop}}(M_Z^2)=0.1166\pm0.0013;\cr
\hbox{\chidof}=& 92/(153-18)\cr}$$
from $ep$ and, from $\nu N$,
$$\eqalign{\lambdav(n_f=4, 3 \;{\rm loop})=&255\pm72\;\mev;
\quad\alpha_s^{3\;{\rm loop}}(M_Z^2)=0.1153\pm0.0063;\cr
\hbox{\chidof}=&0.335/(64-7).\cr} $$
The  errors take
into account  
 experimental errors and 
 higher twist effects  among other estimated 
theoretical errors. 

We can also extrapolate the fit, made originally for the range $3.5\leq Q^2\leq 230\;\gev^2$ 
to higher momenta, obtaining very good reproduction
 of experimental data up to $Q^2\simeq 5000\;\gev^2$. 
>From this, in particular, we can exclude supersymmetric partners (squarks or gluinos)
 for masses up to some 60
\gev. The fitted value of the coupling is now, including the data at high $Q^2$, 
$$\alpha_s^{3\;{\rm loop}}(M_Z^2)=0.1163\pm0.0014 $$}
\centerline{\box0}
\brochureendcover{Typeset with \physmatex}
\pageno=1
\brochureb{\smallsc j. santiago and f. j.  yndur\'ain}{\smallsc improved 
calculation of $F_2$ in 
electroproduction and $xF_3$ in neutrino scattering to nnlo}{1}

\booksection{1 Introduction}
In two recent papers\ref{1,2} (where we send for more details and 
and more extensive introduction and references) 
we have evaluated, and compared to experiment, the 
evolution of the structure function $F_2(x,Q^2)$ in $ep$ 
deep inelastic scattering (DIS) to next to next to leading order in QCD 
perturbation theory (NNLO): 
two loops in the Wilson coefficients, and three loops
 in the anomalous dimensions and strong coupling, $\alpha_s$. 
We used the known values of the two loop Wilson coefficients\ref{3}
 and three loop anomalous dimensions\ref{4} for $n=2,\,4,\,6,\,8$ for the moments 
for quark nonsinglet (NS), singlet (S) and gluon (G) functions, 
 
$$\mu_i(n;Q^2)=\int_0^1\dd x\,x^{n-2}F_i(x,Q^2);\quad i=NS,\,S,\,G.\equn{(1.1)}$$ 
with 
$$F_2(x,Q^2)=F_S(x,Q^2)+F_{NS}(x,Q^2).\equn{(1.2)}$$

The evaluation used the method of Bernstein moments\ref{1,5}. 
In this method one does not compare with experiment the values of 
the $\mu(n,Q^2)$ themselves; these cannot be evaluated reliably due to 
existence, for each value of $Q^2$, of only 
measurements of $F_2(x,Q^2)$ for a limited range of $x$. 
Instead, we compare  theoretical predictions and experimental results for 
 the Bernstein averages,
$$F_{nk}(Q^2)\equiv
\int_0^1\dd x\,p_{nk}(x)F_2(x,Q^2).
\equn{(1.3)}$$
Here the $p_{nk}$ are the (modified) Bernstein polynomials,
$$\eqalign{p_{nk}(x)=&
\dfrac{2\Gammav(n+\tfrac{3}{2})}
{\Gammav(k+\tfrac{1}{2})\Gammav(n-k+1)}x^{2k}(1-x^2)^{n-k}\cr
=&\dfrac{2(n-k)!\Gammav(n+\tfrac{3}{2})}{\Gammav(k+\tfrac{1}{2})\Gammav(n-k+1)}
\sum_{l=0}^{n-k}\dfrac{(-1)^l}{l!(n-k-l)!}x^{2(k+l)};\quad k\leq n.\cr}
\equn{(1.4)}$$
These polynomials  are positive and
 have a single maximum located at 
$$\bar{x}_{nk}=\dfrac{\Gammav(k+1)\Gammav(n+\tfrac{3}{2})}
{\Gammav(k+\tfrac{1}{2})\Gammav(n+2)};$$
 they are concentrated around this point,
 with a spread of 
$$\lap x_{nk}=
\sqrt{\dfrac{k+\tfrac{1}{2}}{n+\tfrac{3}{2}}-
\left[\dfrac{\Gammav(k+1)\Gammav(n+\tfrac{3}{2})}
{\Gammav(k+\tfrac{1}{2})\Gammav(n+2)}\right]^2},$$
 and they are normalized to unity, 
$\int^1_0\dd x\,p_{nk}(x)=1$. Therefore, the integral
$$\int_0^1\dd x\,p_{nk}(x)F_2(x,Q^2)$$
represents an average of the 
function $F_2(x)$ in 
the region $\bar{x}_{nk}-\tfrac{1}{2}\lap x_{nk}\lsim
 x\lsim\bar{x}_{nk}+\tfrac{1}{2}\lap x_{nk}$;
 the 
values of the function  outside this interval contribute  little
 to the integral, as $p_{nk}(x)$ 
decreases to zero very quickly there. 
So, by choosing suitably $n,\,k$ we manage to adjust the region where
 the average is peaked to that in which we have experimental data.  Finally, 
and using the binomial expansion in \equn{(1.3)}, it follows that 
the averages with the $p_{nk}$ of $F_2$ can be obtained in terms of 
its {\sl even} moments:
$$\eqalign{\int_0^1\dd x\,p_{nk}(x)F_2(x)=&
\dfrac{2(n-k)!\Gammav(n+\tfrac{3}{2})}{\Gammav(k+\tfrac{1}{2})\Gammav(n-k+1)}
\sum_{l=0}^{n-k}\dfrac{(-1)^l}{l!(n-k-l)!}\,\mu_{2k+2l+2}.\cr}$$
This is then evaluated theoretically and the result compared to the experimentally evaluated 
averages.

In the present note we improve on this as follows. 
First of all, for $ep$ DIS, we incorporate the values of
 the three loop anomalous dimensions recently available\ref{6} for $n=10,\;12$. 
Secondly, we also evaluate the evolution of the moments 
(Bernstein averages, in fact) for the structure function 
$xF_3$ in $\nu N$ 
DIS (neutrino scattering on an isoscalar target). 
This we can do because the three loop anomalous dimensions for 
$n=1,\dots,13$ have also been evaluated recently\ref{6}. 
So we have here a true NNLO calculation, as opposed to existing 
{\sl estimates} where the value of the anomalous dimensions was 
extrapolated.

The use of more moments means that we can also evaluate more Bernstein averages, 
and these more precisely,  which in turn 
permits an important decrease of the {\sl statistical} errors of the calculation. 
Because we have more moments, we can avoid 
the more suspect, lower $Q^2$ data points, where uncontrollable higher twist 
terms or corrections due to the $c$ quark mass (for example) are largest. So, 
for  $ep$ scattering, the energy range we choose is  
 $3.5\,\gev^2\leq Q^2\leq 230\,\gev$.
 
We can also diminish the {\sl systematic} errors by using the following improvement on the 
previous analysis. We profit from the fact that the structure functions are 
(proportional to) quark densities, hence positive-definite. 
This means that the moments must satisfy 
a set of very powerful determinantal inequalities, to be discussed in \sect~2, 
which force a rigid behaviour of the moments and hence 
avoid spurious minima and spurious systematic deviations. 
This is particularly important for $ep$ scattering, where we have 
eighteen input moments as free parameters.
This is the second of the improvements that allow us a very 
precise determination of the strong coupling. 
We find,

$$\eqalign{\alpha_s^{(\rm 3\;loop)}(M_Z^2)=&
0.1166\pm0.0009\;(\hbox{statistical})\pm0.0010\;(\hbox{systematic})\cr
&\hbox{(for $ep$)},\cr}\equn{(1.7)}$$
and
$$\eqalign{\alpha_s^{(\rm 3\;loop)}(M_Z^2)=&
0.1153\pm0.0041\;(\hbox{statistical only)}\cr
&\hbox{(for $\nu N$)}.\cr}\equn{(1.8)}$$
The results are compatible one with the other, and both with our ``old" result\ref{1}
 for $ep$,
$$\eqalign{\alpha_s^{(\rm 3\;loop)}(M_Z^2)=&
0.1172\pm0.0017\;(\hbox{statistical})\pm0.0017\;(\hbox{systematic})\cr
=&0.1172\pm0.0024\;\hbox{($ep$)},\cr}\equn{(1.9)}$$
but the errors are now substantially smaller.

The quality of the fit is so good that we can consider extending the 
analysis to higher values of $Q^2$, where there are not sufficient experimental points to 
get Bernstein averages with only {\sl interpolation}, so we have to resort to 
{\sl extrapolation} to evaluate the integrals. 
This allows us to include data for $ep$ scattering
 with momenta as high as $Q^2\simeq 5000\;\gev^2$. 
We get two important results: i) If we include the new points in the fit, the value of 
$\alpha_s$ obtained only changes to 
$$\alpha_s^{(\rm 3\;loop)}(M_Z^2)=
0.1163\pm0.0010\quad\hbox{(statistical error only)};$$
 or we can simply extrapolate the theoretical fit obtained in the 
restricted range $3 - 230\;\gev^2$. Then we find perfect agreement with the result obtained fitting 
the full range. 
These results show the stability and reliability of the calculations, 
as well as their suitability for high energy extrapolation. 

As a byproduct of this last analysis we can extend 
the results of ref.~2 to give bounds 
$$m_{\tilde{q}},\,m_{\tilde{g}}\geq\,60\;\gev$$
 for the masses of squarks or gluinos.
\vfill\eject

\booksection{2 Bernstein Averages and Positivity Inequalities}
\booksubsection{2.1 $ep$ Scattering}
We consider as input parameters, to be fitted through the procedure to be 
described, the eighteen moments at a given reference momentum $Q_0^2$,
$$\mu_i(n,Q_0^2),\quad n=2,\,4,\,6,\,8,\,10,\,12;
\quad i=NS,\,S,\,G.\equn{(2.1)}$$
Actually, the momentum sum rule allows us to obtain $\mu_G(2,Q_0^2)$ 
in terms of $\mu_S(2,Q_0^2)$, so we have only seventeen 
independent momenta; but, if we also fit the QCD coupling 
we have also $\Lambdav$, so we still end up with 18 parameters.
We evolve the values of these moments with the QCD
 evolution equations (for the details, see ref.~1) so we 
get the moments to all values of $Q^2$:
$$\mu_i(n,Q^2),\quad n=2,\,4,\,6,\,8,\,10,\,12;
\quad i=NS,\,S,\,G.\equn{(2.2)}$$
With these momenta we form the theoretical 
Bernstein averages, which are then fitted to their experimental values. 
However, not any arbitrary input $\mu_i(n,Q_0^2)$ are acceptable; 
they have to be such that, not only they fit the experimental data, but, 
{\sl at all $Q^2$}, they have to be the moments of 
positive-definite functions. 

It is possible to implement this restriction by 
imposing a set of determinantal inequalities, which are the necessary and 
sufficient conditions for the $\mu_i(n,Q^2)$ to be the moments of 
positive quark and gluon distribution functions. 
The details may be found in the Appendix, or in refs.~7,8. 
The inequalities may be 
written as
$$\Delta_\nu[\mu(n,Q^2)]\geq 0\quad\nu=1,\,2,\dots.\equn{(2.2)}$$
where the $\Delta_\nu$ are determinants formed with the moments. 
In our case, if we define $M_k(Q^2)\equiv \mu(2k+2,Q^2)$, 
$k=0,\dots,5$,  
(for any of the $NS,\,S,\,G$ cases) we can write (2.2) as 
the explicit  inequalities that guarantee that 
the determinants
$$\det\pmatrix{M_1&M_2&M_3\cr
M_2&M_3&M_4\cr
M_3&M_4&M_5\cr},$$

$$\det\pmatrix{M_0-M_1&M_1-M_2&M_2-M_3\cr
M_1-M_2&M_2-M_3&M_3-M_4\cr
M_2-M_3&M_3-M_4&M_4-M_5\cr}$$
and all their principal minors are {\sl positive}.
Corresponding inequalities hold for the case when an odd number of moments is known.

In principle, the inequalities (2.2) should be strictly 
satisfied, i.e., we should impose them on the fits. 
However, one should take into account that the calculation 
is not exact: higher twist, or {\sl N}NNLO effects are 
not included, and one should therefore allow a certain laxity. 
What we have done is to add to the chi-squared a 
penalty if some of the determinants are not positive. 
The penalty is obtained by multiplying the value of the negative
 determinants by $10^7$, and adding all these contributions to the chi-squared. 
This is done for {\sl all} the values of $Q^2$. 
In the actual fits, the sum total of the values of all the negative 
determinants is less than $10^{-6}$, a quantity 
well within the tolerance due to the 
 estimated errors stemming from  higher order effects.

\booksubsection{2.2 $\nu N$ Scattering}
The only difference in this case is that only {\sl odd} moments are known now:
$$\mu(n,Q^2)=\int_0^1\dd x\, x^{n-2} xF_3(x,Q^2),\quad n=1,\,3,\,5,\,7,\,9,\,11,\,13.
\equn{(2.3)}$$
Therefore, to form the quantities $M_i$ (see the Appendix) one has to 
consider $F_3$ to be the structure function. 
The rest is as for the $ep$ case, except that we have an odd 
number of moments now.
\vfill
\eject
\booksection{3 Computations and Results}
\booksubsection{3.1 $ep$ Scattering}
With the available values for the anomalous dimensions 
we can use the first six (even) moments. 
When evaluating the experimental input, we have to calculate the 
average of the experimental $F_2$ with Bernstein polynomials. 
In order not to rely on extrapolations, we follow the criterion stated in 
ref.~1 (where we send for 
details): we only include an average with $p_{nk}$ {\sl if we have experimental points 
covering the whole region}, 
$[\bar{x}_{nk}-\Delta\bar{x}_{nk},\bar{x}_{nk}+\Delta\bar{x}_{nk}]$. 
For $ep$ scattering this means that we can use the averages
$$\eqalign{F^{\rm exp}_{10}(Q^2),\quad F^{\rm exp}_{20}(Q^2),\quad F^{\rm exp}_{30}(Q^2),\cr
F^{\rm exp}_{21}(Q^2),\quad F^{\rm exp}_{31}(Q^2),\quad F^{\rm exp}_{32}(Q^2),\cr
F^{\rm exp}_{41}(Q^2),\quad F^{\rm exp}_{42}(Q^2),\cr
F^{\rm exp}_{51}(Q^2),\quad F^{\rm exp}_{52}(Q^2);\cr}
\equn{(3.1)}$$
but of course not for each value of $Q^2$ we have enough 
experimental information to evaluate all the moments. 
Altogether, we have 153 experimental averages.\fnote{A point to 
take into account is that not all the averages are independent; 
because, for each  $Q^2$ we have only six moments, of the 
12 possible averages only 6 are independent. This can be taken into account 
by renormalizing the chi-squared scaling it to the number of independent 
experimental values; see again ref.~1 for details. 
This scaling has been taken into account, 
so the \chidof presented here refers to the truly independent degrees of freedom.}
The values of the $Q^2$ range from 3.5 $\gev^2$ to 230 $\gev^2$.
The experimental input comes from ref.~9 for $ep$ and 
ref.~10 for $\nu N$. 
The result of the fit may be seen in \fig~1.
\topinsert{
\setbox3=\vbox{\hsize 11.6truecm
\setbox0=\vbox{\epsfxsize 8.truecm\epsfbox{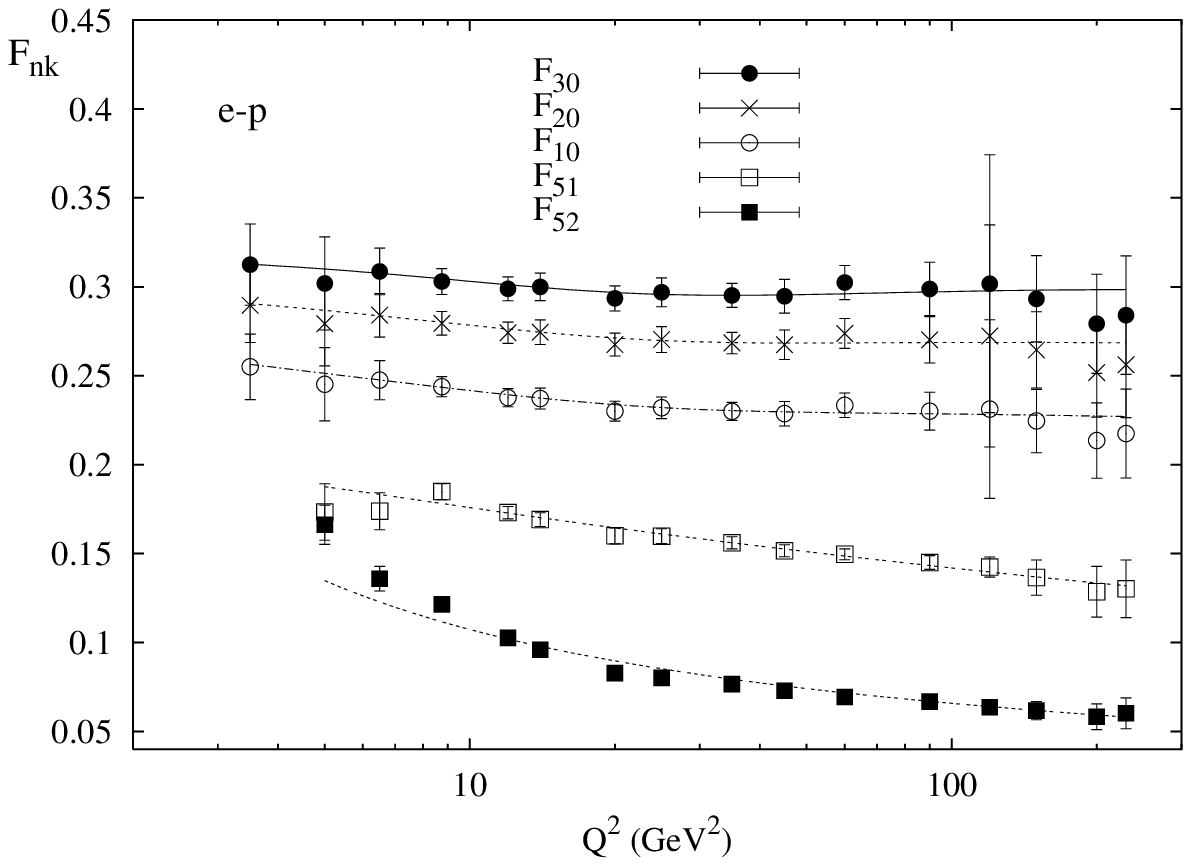}} 
\centerline{{\box0}}
\medskip
\centerline{\box 3}
\setbox7=\vbox{\hsize 11.6truecm
\setbox5=\vbox{\epsfxsize 8.truecm\epsfbox{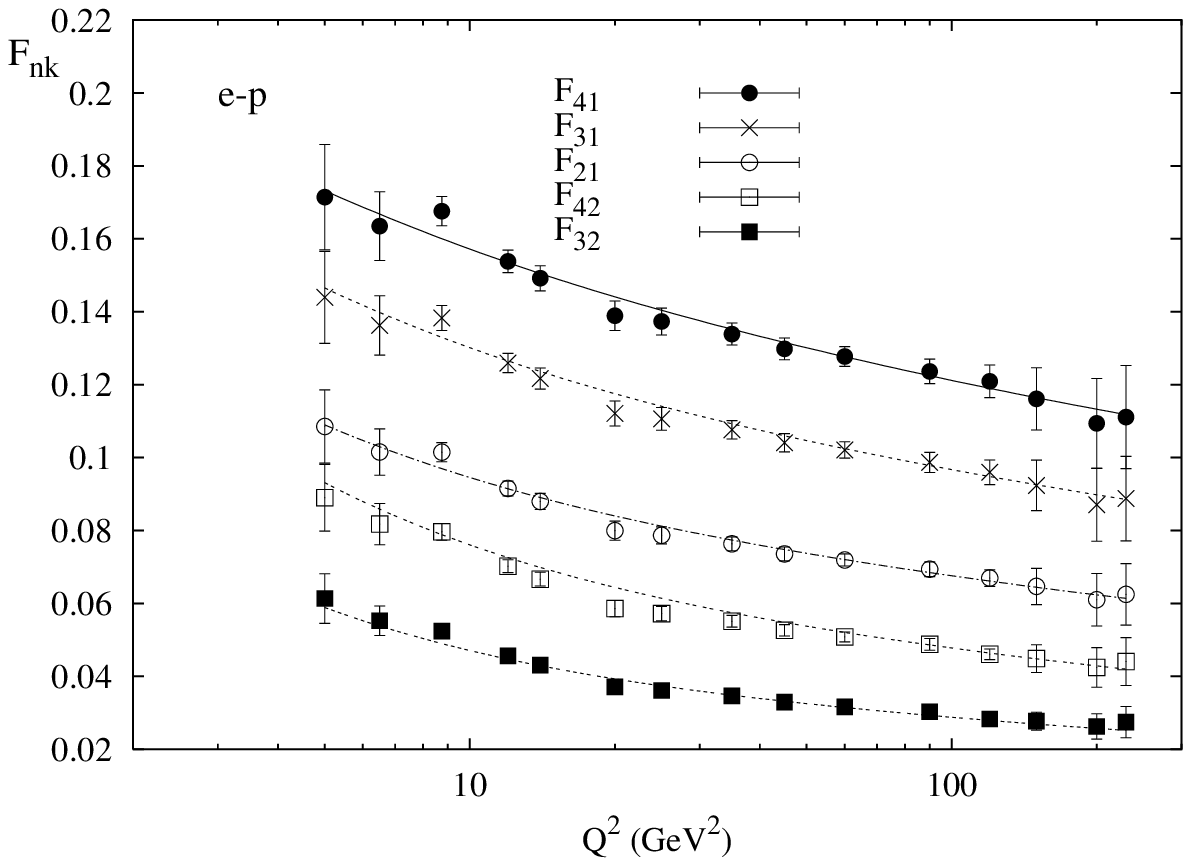}}
\setbox6=\vbox{\hsize 11.5truecm\captiontype\figurasc{Figure 1. }{NNLO fit 
to  Bernstein averages of  $F_2$; TMC  included.}\hb
\vskip.1cm} 
\centerline{{\box5}}
\medskip
\centerline{\box6}}
\centerline{\box7}}
\centerline{\box3}
}\endinsert

Before presenting the numerical results, a few words have to be said both 
on the obtention of the experimental input and on the theoretical 
evaluation. As to the first, we remark that 
the calculation of an average such as (1.3),
$$F_{nk}^{\rm exp}(Q^2)\equiv
\int_0^1\dd x\,p_{nk}(x)F_2^{\rm exp}(x,Q^2),
\equn{(3.2)}$$
 requires 
some interpolation and extrapolation of $F_2^{\rm(exp)}(x,Q^2)$. To do so 
we have used two different methods. We separate $F_2$ into a singlet 
and a nonsinglet part, writing
$$F_2=F_S+F_{NS}.$$
In the first method we use, 
for each 
value of $Q^2$ independently, a phenomenological 
expression for the $F$,
$$\eqalign{F_S^{\rm phen.}(x)=&(Ax^{-0.44}+C)(1-x)^\nu,\cr
F_{NS}^{\rm phen.}(x)=&B x^{0.5}(1-x)^\mu.\cr}
\equn{(3.3)}$$
We  emphasize that, 
in the present paper, (3.3) is to be considered as only a convenient 
interpolation of the data to allow calculation of the 
integrals (3.2). In fact, 
the parameters $A,\,B,\,C,\,\nu,\,\mu$ are taken to be 
totally free, and assumed to be {\sl uncorrelated}  
for different values of $Q^2$. Thus, no theoretical 
bias is induced in the $Q^2$ dependence. 

The second method we consider is to use 
a {\sl polynomial} parametrization of data.  Then the integral involved in the  Bernstein average 
is evaluated with the help of the parametric expression, for each value of $Q^2$. 
A third method will also be used (for the nonsinglet piece); 
it will be discussed when we deal with $\nu N$ scattering. 
As shown below, our results depend very little 
on the method of interpolation.

We consider the first method to be the cleanest one; the results 
found using the polynomial parametrization are presented mostly 
to show the insensitivity of the 
evaluation to the method of obtaining the ``experimental" 
averages $F^{\rm(exp)}_{nk}(Q^2)$. 
By varying $F^{\rm exp}_2$ within its experimental 
error bars (including {\sl experimental} systematic errors) 
we obtain the averages, and their errors. 
The fit to these provides the values of our parameters and what we call our 
{\sl statistical} error. 

The {\sl systematic} error of our fit is really the resultant of 
the estimated theoretical errors, 
which we briefly discuss here: full details 
may be found in ref.~1. 

As stated before, we take as parameters of our fit the values 
of the moments at a reference momentum, $\mu_i(n,Q^2_0)$. 
For the fit we choose $Q_0^2=8.75\;\gev^2$. We may consider changing this 
to another value: the modification our results 
will suffer will be a measure of  higher order effects. 
Other higher order effects are the {\sl N}NNLO corrections,
 that we  estimate both as in ref.~1, and by using the four loop 
expression\ref{11} for $\alpha_s$; higher twist corrections, that we incorporate 
phenomenologically by adding a contribution
$$\mu_{NS}^{HT}(n,Q^2)=n(a\Lambdav^2/Q^2)\mu_{NS}(n,Q^2)
\equn{(3.4)}$$ 
with $a$ an unknown parameter, to be fitted, and expected 
to be of order unity. 
Next, we take into account target mass corrections (TMC), to order 
$m_p^2/Q^2$ by writing (see again ref.~1 for details) 
$$
\mu^{TMC}_{NS}(n,Q^2)=\mu_{NS}(n,Q^2)+\dfrac{n(n-1)}{n+2}
\dfrac{m_p^2}{Q^2}\mu_{NS}(n+2,Q^2).
$$
It should be noted that the TMCs for the singlet are negligible 
compared to those of the NS, which is why we only take 
into account the last.
We remark that TMC are incorporated into our best fit, but 
higher twist corrections or 
NNNLO ones are only used to estimate the theoretical uncertainty.

Finally, we consider the 
matter of the number of quark flavours. 
The only quark threshold we cross is that of the $b$ quark. 
We will be working with the moments $\mu(n,Q^2)$, so we 
have only one momentum variable. We then split the $Q^2$
 range into the following two intervals:
$$Q^2\lsim m_b^2\;({\rm I});\quad m_b^2\lsim Q^2\;{(\rm II)}.$$ 
Then, in region (I) we take $n_f=4$ and in region 
(II), $n_f=5$. The matching will be carried 
only for the coupling constant, following the standard prescription: to NNLO,\ref{12}
$$\beta_0^{n_f+1}\log\dfrac{\Lambdav^2(n_f+1)}{\Lambdav^2(n_f)}=
(\beta_0^{n_f+1}-\beta_0^{n_f})L_h+\delta_{NLO}+\delta_{NNLO}
$$
where
$$\eqalign{\delta_{NLO}=&
(b_1^{n_f+1}- b_1^{n_f})\log L_h-b_1^{n_f+1}
\log\dfrac{\beta_0^{n_f+1}}{\beta_0^{n_f}},\cr
\delta_{NNLO}=&\dfrac{1}{\beta_0^{n_f}L_h}
\left[(b_1^{n_f+1}- b_1^{n_f})b_1^{n_f}\log L_h+
(b_1^{n_f+1})^2-(b_1^{n_f})^2+b_2^{n_f}-b_2^{n_f+1}+\tfrac{7}{24}\right].\cr}
$$
Here,
$$L_h=\log\left[m^2(n_f+1)/\Lambdav^2(n_f)\right],\quad b_i=\beta_i/\beta_0$$
and $m(n_f+1)$ is the pole mass of the $(n_f+1)$th quark. 
For its value we take that of ref.~13.

It is important to realize that this method only provides an 
approximated evaluation of threshold effects, which are 
somewhat distorted in the neighbourhood of $Q^2=m_b^2$. 
To estimate the corresponding error we use
an alternate method, which  is to avoid the region $Q^2\simeq m_b^2$, i.e., to 
remove from the fit the corresponding experimental points.

We  present in Table 1 a compilation of the results obtained
 with our calculations at LO, NLO and 
NNLO, with TMCs taken into account; the fit to the data itself is shown,
 for the NNLO calculation (with TMC) in \fig~1. 
Only {\sl statistical} errors are shown in Table 1;
systematic (theoretical) errors will  be discussed below. We note that 
the errors given in Table 1 are ``renormalized" 
to take into account the effective number of independent 
experimental points, 
as discussed before.
\bigskip  

\setbox0=\vbox{
\setbox1=\vbox{\offinterlineskip\hrule
\halign{
&\vrule#&\strut\hfil#\hfil&\vrule#&\strut\hfil#\hfil&
\quad\vrule\quad#&\strut\quad#\quad&\quad\vrule#&\strut\quad#\cr
 height2mm&\omit&&\omit&&\omit&&\omit&\cr 
&\phantom{l}Order\phantom{l}&&\phantom{l}$\Lambdav(n_f=4)$&
&$\alpha_s(M_Z^2)$\kern.3em&&\chidof& \cr
 height1mm&\omit&&\omit&&\omit&&\omit&\cr
\noalign{\hrule} 
height1mm&\omit&&\omit&&\omit&&\omit&\cr
&LO&&\phantom{\big|}$ 226\pm34$&&$0.131\pm0.003$&&$130/(153-18)\;$& \cr
\noalign{\hrule} 
height1mm&\omit&&\omit&&\omit&&\omit&\cr
&NLO&&\phantom{\big|}$ 279\pm19$&&$0.1155\pm0.0014$&&$97/(153-18)\;$& \cr
\noalign{\hrule} 
height1mm&\omit&&\omit&&\omit&&\omit&\cr
&NNLO&&\phantom{\big|}$ 274\pm13$&&$0.1166\pm0.0009$&&$92/(153-18)=0.68\;$& \cr
 height1mm&\omit&&\omit&&\omit&&\omit&\cr
\noalign{\hrule}}
\vskip.05cm}
\centerline{\box1}
\smallskip
\centerline{\petit Table 1}
\centerrule{6cm}
\medskip}
\box0
The NLO corrections are very clearly seen in the 
fit: the 
\chidof\ decreases substantially  
when including these. It also decreases when adding NNLO corrections, 
but the fit is so good already at LO  
 that there is very little room for improvement. 
However the improvement is seen, particularly in that
 including NNLO corrections leads to 
 a noticeable gain both in the 
quality of the determination of the coupling, and in the {\sl stability} of the fits. 
Also the improvement in the quality of the fit 
with respect to that of refs.~1,2 is apparent; not only the errors 
are smaller, but the \chidof\ has also substantially decreased, from $79/(102-12)=0.88$ 
to the value $92/(153-18)=0.68$ we get now. In fact, the 
precision improves also to LO and NLO, thus 
emphasizing that the improvement, in particular the  
important gain in stability, is due,to a large extent, to 
the use of the positivity constraints that prevent the appearance of spurious minima.

The estimated systematic errors, originating  from various sources,
 are  shown for the NNLO case  
in Table 2. It is interesting to remark that the use of the 
positivity constraints also improves the systematic, 
theoretical errors, which are now smaller than in the 
estimates of refs.~1,2, with the exception of the 
one due to estimated NNNLO effects. But this is because
 we also include here another source of error, namely, that
 due to cutting the expansion of $\alpha_s$ to three loops. We have,

\setbox0=\vbox{
\medskip
\setbox1=\vbox{\offinterlineskip\hrule
\halign{
&\vrule#&\strut\hfil#\hfil&\vrule#&\strut\hfil#\hfil&
\quad\vrule\quad#&\strut\quad#\quad&\quad\vrule#&\strut\quad#\cr
 height2mm&\omit&&\omit&&\omit&&\omit&\cr 
&\phantom{l}Source of error\phantom{l}&&\phantom{l}$\Lambdav(n_f=4;\hbox{3 loop})$&
&\phantom{l}$\lap\Lambdav(n_f=4;\hbox{3 loop})$&&$\lap\alpha_s(M_Z^2)$\kern.3em& \cr
 height1mm&\omit&&\omit&&\omit&&\omit&\cr
\noalign{\hrule} 
height1mm&\omit&&\omit&&\omit&&\omit&\cr
&No TMC&&\phantom{\big|}$279$&&\hfil$5$\hfil&&$0.0004$& \cr
\noalign{\hrule} 
height1mm&\omit&&\omit&&\omit&&\omit&\cr
&\phantom{\big|}Interpolation 
\phantom{\big|}&&\phantom{\big|}$ 279$&&\hfil$5$\hfil&&$0.0004$& \cr
\noalign{\hrule} 
height1mm&\omit&&\omit&&\omit&&\omit&\cr
&HT&&\phantom{\big|}$268$&&\hfil$6$\hfil&&$0.0004$& \cr
\noalign{\hrule} 
height1mm&\omit&&\omit&&\omit&&\omit&\cr
&Quark mass effect&&269&&\hfil$5$\hfil&&$0.0004$& \cr
\noalign{\hrule}
height1mm&\omit&&\omit&&\omit&&\omit&\cr
&$Q_0^2$ to $12\;\gev^2$&&\phantom{\big|}$279$&&\hfil$5$\hfil&&$0.0004$& \cr
\noalign{\hrule}
height1mm&\omit&&\omit&&\omit&&\omit&\cr
&{\sl N}NNLO&&\phantom{\big|}$265$&&\hfil$10$\hfil&&$0.0006$& \cr
 height1mm&\omit&&\omit&&\omit&&\omit&\cr
\noalign{\hrule}}
\vskip.05cm}
\centerline{\box1}
\smallskip
\centerline{\petit Table 2}
\centerrule{6cm}
\medskip}
\box0
\noindent Let us comment on the meaning of the different entries. 
No TMC means that we have {\sl not} taken 
target mass corrections into account. The 
corresponding  
error is {\sl not} included when evaluating the overall theoretical error 
because, since we take into account TMC in 
our central value, the error 
would be of order $TMC^2$, or $\alpha_s\times TMC$, 
quite negligible. 
 ``Interpolation" means that we have used the polynomial interpolation   
to calculate the Bernstein averages.  
 HT means that we have 
taken into account the presence of higher 
twist by adding 
a contribution like  (3.4). The fitted value of the 
phenomenological parameter  $a$ is $a=-0.003\pm0.002$. 
 ``Quark mass effect" means that we have 
cut off the $b$ quark threshold, as discussed; 
the error in Table 2 takes into account 
also the variations due to the error in the 
$m_b$ mass, taken (with its value) from ref.~13. ``$Q_0^2$ to $12\,\gev^2$" means 
that we take the input moments to be defined at this 
value of the momentum, $\mu_i(n,Q_0^2=12\,\gev^2)$, $i=S,\,G,\,NS$.
 Finally, {\sl N}NNLO means that we have fitted with 
the four loop  formula for $\alpha_s$.

Composing quadratically systematic (theoretical) and statistical (experimental)  
errors we find the best result for the QCD coupling, 
$$\eqalign{\Lambdav(n_f=4,\,\hbox{3 loop})=&274\pm13\;(\hbox{stat.})\pm15\;(\hbox{syst.})
=274\pm20\;\mev;\cr 
 \alpha_s^{(\rm 3\, loop)}(M_Z)=&0.1166\pm0.0009\;(\hbox{stat.})\pm0.0010\;(\hbox{syst.})
=0.1166\pm0.0013.\cr} 
\equn{(3.5)}$$
The corresponding central value for $\lambdav(n_f=5,\,\hbox{3 loop})$ 
is of $193\,\mev$. 

It is to be noted that composing the ``theoretical" 
errors 
as if they were independent leads to a certain amount of double-counting. Thus, the 
results should be independent of the value 
of $Q_0^2$ if the calculation was to all orders, so the 
two last errors in Table 2 are connected, and our estimate of the ``systematic" 
(theoretic) error is really an overestimate.

An important outcome of our 
calculations are the values of the moments of the  quark and 
gluon densities, which should be useful as constraints for 
model-builders. 
These calculations we have performed in two ways. 
First, we can take directly the results of the fits. 
Secondly, we can impose with greater strength the 
positivity constraints at this initial value, 
$Q^2=Q^2_0$ by weighting the determinantal inequalities that ensure 
positivity more heavily there, so that 
the corresponding values of the moments are more reliable. 
In both cases, the values of all the moments for the NS case, and of the lowest $n$ 
(up to $n=8$) for the quark singlet and gluon moments are essentially unchanged. 
Also, the values of $\lambdav$ is left essentially unchanged, another 
proof of the stability of our results. 
The corresponding values of the moments are then, 
 at our starting values of $Q^2=Q^2_0$:
\medskip

$$\matrix{Q_0^2=8.75 &\quad Q_0^2=12\cr
\mu_{NS}(2)=0.126\pm0.006&\quad0.122\pm0.005\cr
\mu_{NS}(4)=0.0161\pm0.0011&\quad0.0150\pm0.0010\cr
\mu_{NS}(6)=0.0036\pm0.0005&\quad0.0032\pm0.0008\cr
\mu_{NS}(8)=0.0011\pm0.0005&\quad0.0009\pm0.0004\cr
\mu_{NS}(10)=0.00041\pm0.00017&\quad0.00026\pm0.00010\cr
\mu_{NS}(12)=0.00020\pm0.00011&\quad0.00011\pm0.00006,\cr}$$
\medskip

$$\matrix{Q_0^2=8.75 &\quad Q_0^2=12\cr
\mu_S(2)=0.0521\pm0.0020\phantom{XY}&\qquad0.0519\pm0.0015\cr
\mu_S(4)=(0.62\pm0.06)\times10^{-4}&\qquad(0.77\pm0.06)\times10^{-4}\cr
\mu_S(6)=(0.24\pm0.08)\times10^{-4}&\qquad(0.34\pm0.08)\times10^{-4}\cr
\mu_S(8)=(0.10\pm0.05)\times10^{-4}&\qquad(0.23\pm0.07)\times10^{-4}\cr
\mu_S(10)=(0.39\pm0.15)\times10^{-5}&\qquad(0.10\pm0.05)\times10^{-4}\cr
\mu_S(12)=(0.16\pm0.07)\times10^{-5}&\qquad(0.41\pm0.20)\times10^{-5}\cr}$$
and

$$\matrix{Q_0^2=8.75 &\quad Q_0^2=12\cr
\mu_G(2)=0.208\pm0.006\phantom{XYZ}&\qquad0.209\pm0.004\cr
\mu_G(4)=(0.94\pm0.008)\times10^{-3}&\qquad(0.85\pm0.08)\times10^{-3}\cr
\mu_G(6)=(0.24\pm0.06)\times10^{-3}&\qquad(0.27\pm0.05)\times10^{-3}\cr
\mu_G(8)=(0.78\pm0.28)\times10^{-4}&\qquad(0.13\pm0.03)\times10^{-3}\cr
\mu_G(10)=(0.33\pm0.12)\times10^{-4}&\qquad(0.50\pm0.20)\times10^{-4}\cr
\mu_G(12)=(0.10\pm0.06)\times10^{-4}&\qquad(0.20\pm0.10)\times10^{-4}\cr}$$

As discussed, the results are  fairly stable; but we should emphasize that they do {\sl not} include 
systematic errors, which are expected to be largest for large $n$, especially  
for the quark singlet and gluon moments.

If we do not overweight the 
reference moments we would have got the following values:
$$\matrix{Q_0^2=8.75 &\quad Q_0^2=12\cr
\mu_{NS}(2)=0.126\pm0.006&\quad0.122\pm0.005\cr
\mu_{NS}(4)=0.0160\pm0.0010&\quad0.0150\pm0.0010\cr
\mu_{NS}(6)=0.0036\pm0.0005&\quad0.0032\pm0.0010\cr
\mu_{NS}(8)=0.0010\pm0.0005&\quad0.0009\pm0.0004\cr
\mu_{NS}(10)=0.00034\pm0.00011&\quad0.00026\pm0.00009\cr
\mu_{NS}(12)=0.00017\pm0.00009&\quad0.00012\pm0.00007,\cr}$$
\medskip
$$\matrix{Q_0^2=8.75 &\quad Q_0^2=12\cr
\mu_S(2)=0.0522\pm0.0015\phantom{XY}&\qquad0.0519\pm0.0008\cr
\mu_S(4)=(0.58\pm0.05)\times10^{-4}&\qquad(0.93\pm0.03)\times10^{-4}\cr
\mu_S(6)=(0.27\pm0.08)\times10^{-4}&\qquad(0.54\pm0.02)\times10^{-4}\cr
\mu_S(8)=(0.14\pm0.07)\times10^{-4}&\qquad(0.41\pm0.03)\times10^{-4}\cr
\mu_S(10)=(0.41\pm0.12)\times10^{-5}&\qquad(0.94\pm0.10)\times10^{-5}\cr
\mu_S(12)=(0.22\pm0.10)\times10^{-5}&\qquad(0.55\pm0.09)\times10^{-5}\cr}$$
and
$$\matrix{Q_0^2=8.75 &\quad Q_0^2=12\cr
\mu_G(2)=0.208\pm0.004\phantom{XYZ}&\qquad0.210\pm0.003\cr
\mu_G(4)=(0.86\pm0.004)\times10^{-3}&\qquad(0.87\pm0.04)\times10^{-3}\cr
\mu_G(6)=(0.29\pm0.06)\times10^{-3}&\qquad(0.269\pm0.013)\times10^{-3}\cr
\mu_G(8)=(0.13\pm0.08)\times10^{-3}&\qquad(0.95\pm0.11)\times10^{-4}\cr
\mu_G(10)=(0.74\pm0.08)\times10^{-4}&\qquad(0.53\pm0.07)\times10^{-4}\cr
\mu_G(12)=(0.24\pm0.09)\times10^{-4}&\qquad(0.10\pm0.05)\times10^{-4}\cr}$$
The difference with those reported above may be taken as a measure of 
the systematic errors involved in the calculations of the reference {\sl moments}.

\booksubsection{3.2 $\nu N$ Scattering}
The treatment of $xF_3$ in $\nu N$ scattering is essentially similar to 
that of the nonsinglet part of $F_2$ in $ep$ scattering, except 
for the modifications indicated, which are due to our having now only {\sl odd} 
moments. We have now 64 Bernstein averages that can be calculated 
experimentally, and the $Q^2$ range is from 7.9 to 125.9 $\gev^2$.

We do not have now the momentum sum rule, but the Gross--Llewellyn~Smith one, which 
we implement. 
The results of the fit are shown graphically in \fig~2.

\topinsert{
\setbox0=\vbox{\epsfxsize 8.truecm\epsfbox{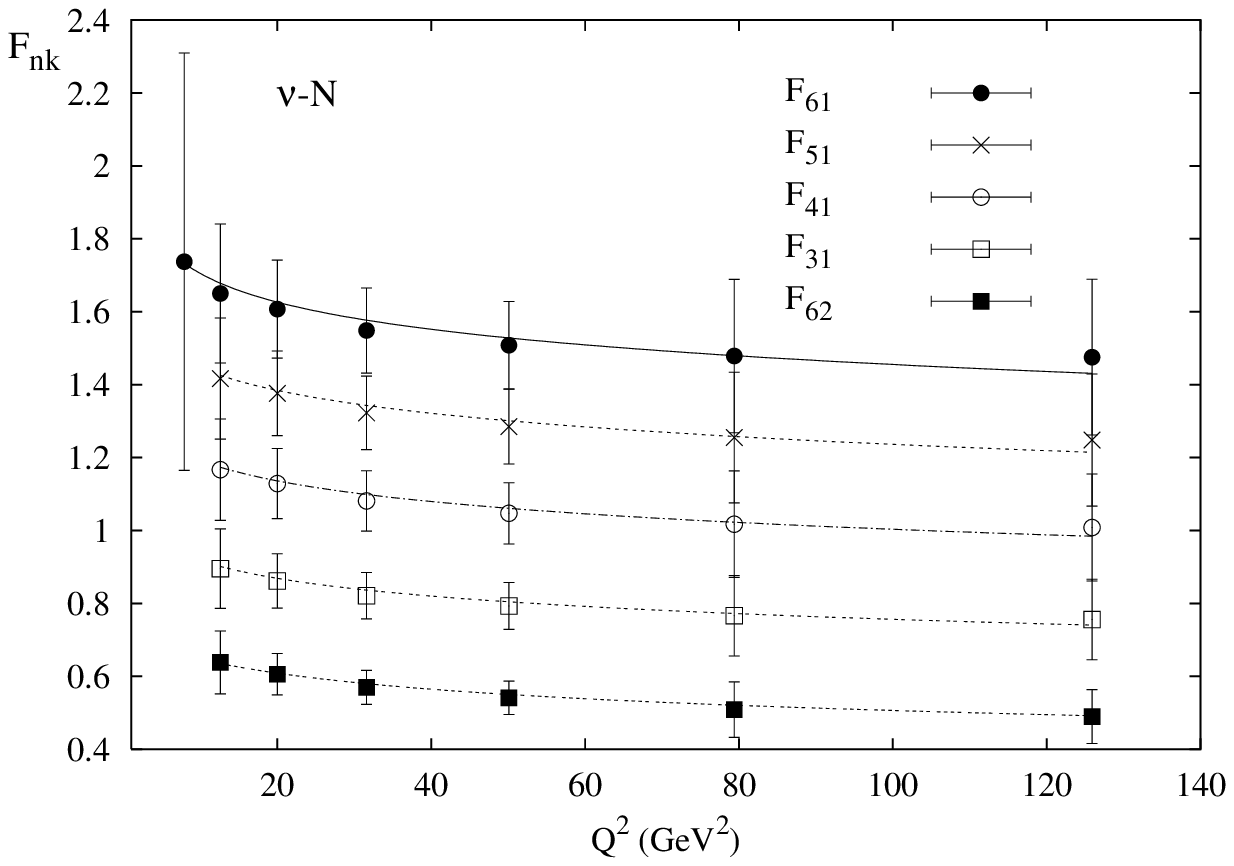}} 
\centerline{{\box0}}

\setbox3=\vbox{\hsize 12.2truecm
\setbox0=\vbox{\epsfxsize 8.truecm\epsfbox{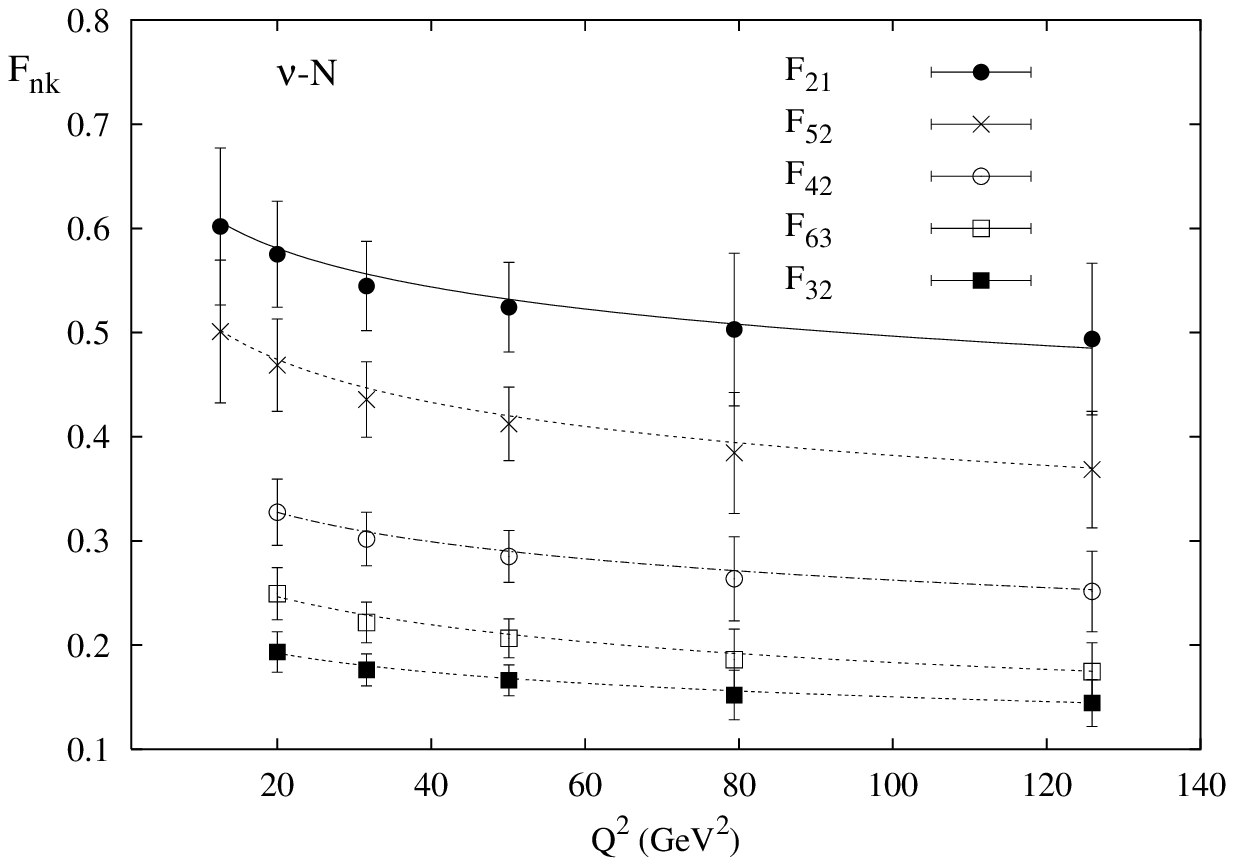}}
\setbox1=\vbox{\hsize 10.5truecm\captiontype\figurasc{Figure 2. }{Fit 
to $xF_3$ with Bernstein averages.}\hb
\vskip.1cm} 
\centerline{{\box0}}
\medskip
\centerline{\box1}}
\centerline{\box 3}
\medskip
}\endinsert
As for the numerical results, the estimated systematic
 errors are as for $ep$ scattering; but, to test the 
dependence of the results on the parametrizations used to get the experimental 
Bernstein averages, we have also calculated replacing the interpolating formula, 
that we now take as
$$xF_3^{\rm phen.}(x)=B x^{\rho}(1-x)^\mu$$
($\rho,\,\mu\,B$ freee parameters) 
by a cubic interpolation, requiring also the 
vanishing of the structure function at both $x=0,\, 1$. 
The results are given in the following Tables~3,4.

\setbox0=\vbox{
\setbox1=\vbox{\offinterlineskip\hrule
\halign{
&\vrule#&\strut\hfil#\hfil&\vrule#&\strut\hfil#\hfil&
\quad\vrule\quad#&\strut\quad#\quad&\quad\vrule#&\strut\quad#\cr
 height2mm&\omit&&\omit&&\omit&&\omit&\cr 
&\phantom{l}Order\phantom{l}&&\phantom{l}$\Lambdav(n_f=4)$&
&$\alpha_s(M_Z^2)$\kern.3em&&\chidof& \cr
 height1mm&\omit&&\omit&&\omit&&\omit&\cr
\noalign{\hrule} 
height1mm&\omit&&\omit&&\omit&&\omit&\cr
&LO&&\phantom{\big|}$ 217\pm78$&&$0.130\pm0.006$&&$0.011\;$& \cr
\noalign{\hrule} 
height1mm&\omit&&\omit&&\omit&&\omit&\cr
&NLO&&\phantom{\big|}$ 281\pm57$&&$0.116\pm0.004$&&$0.008\;$& \cr
\noalign{\hrule} 
height1mm&\omit&&\omit&&\omit&&\omit&\cr
&NNLO&&\phantom{\big|}$ 255\pm55$&&$0.1153\pm0.004$&&$0.38/(64-7)=0.007\;$& \cr
 height1mm&\omit&&\omit&&\omit&&\omit&\cr
\noalign{\hrule}}
\vskip.05cm}
\centerline{\box1}
\smallskip
\centerline{\petit Table 3}
\centerrule{6cm}
\bigskip}
\box0
The \chidof\ is minute, the reason for which is obvious by looking at \fig~2: clearly, 
the experimental errors are much overestimated. 
Nevertheless, it is satisfactory that the \chidof\ 
of the fits decreases from LO to NLO, and even from this to NNLO.
The  systematic errors
 are  shown for the NNLO case  
in Table 4:

\setbox0=\vbox{
\medskip
\setbox1=\vbox{\offinterlineskip\hrule
\halign{
&\vrule#&\strut\hfil#\hfil&\vrule#&\strut\hfil#\hfil&
\quad\vrule\quad#&\strut\quad#\quad&\quad\vrule#&\strut\quad#\cr
 height2mm&\omit&&\omit&&\omit&&\omit&\cr 
&\phantom{l}Source of error\phantom{l}&&\phantom{l}$\Lambdav(n_f=4;\hbox{3 loop})$&
&\phantom{l}$\lap\Lambdav(n_f=4;\hbox{3 loop})$&&$\lap\alpha_s(M_Z^2)$\kern.3em& \cr
 height1mm&\omit&&\omit&&\omit&&\omit&\cr
\noalign{\hrule} 
height1mm&\omit&&\omit&&\omit&&\omit&\cr
&No TMC&&\phantom{\big|}$298$&&\hfil$43$\hfil&&$0.0030$& \cr
\noalign{\hrule} 
height1mm&\omit&&\omit&&\omit&&\omit&\cr
&\phantom{\big|}Cubic interpol. 
\phantom{\big|}&&\phantom{\big|}$ 269$&&\hfil$14$\hfil&&$0.0010$& \cr
\noalign{\hrule} 
height1mm&\omit&&\omit&&\omit&&\omit&\cr
&HT&&\phantom{\big|}$270$&&\hfil$15$\hfil&&$0.0010$& \cr
\noalign{\hrule} 
height1mm&\omit&&\omit&&\omit&&\omit&\cr
&Quark mass effect&&240&&\hfil$15$\hfil&&$0.0010$& \cr
\noalign{\hrule}
height1mm&\omit&&\omit&&\omit&&\omit&\cr
&$Q_0^2$ to $12\;\gev^2$&&\phantom{\big|}$263$&&\hfil$8$\hfil&&$0.0005$& \cr
\noalign{\hrule}
height1mm&\omit&&\omit&&\omit&&\omit&\cr
&{\sl N}NNLO&&\phantom{\big|}$209$&&\hfil$46$\hfil&&$0.0037$& \cr
 height1mm&\omit&&\omit&&\omit&&\omit&\cr
\noalign{\hrule}}
\vskip.05cm}
\centerline{\box1}
\smallskip
\centerline{\petit Table 4}
\centerrule{6cm}
\medskip}
\box0
For the calculation with higher twist contributions, the 
value of the phenomenological parameter $a$ is $a=-0.14\pm1.6$, 
compatible (within the very large errors) with that found for $ep$.

Composing errors as for $ep$ we find,
$$\eqalign{\Lambdav(n_f=4,\,\hbox{3 loop})=&255\pm55\;(\hbox{stat.})\pm46\;(\hbox{syst.})
=255\pm72\;\mev;\cr 
 \alpha_s^{(\rm 3\, loop)}(M_Z)=&0.1153\pm0.0040\;(\hbox{stat.})\pm0.0061\;(\hbox{syst.})
=0.1153\pm0.0063.\cr} 
\equn{(3.6)}$$

The values of the input moments are
$$\matrix{Q_0^2=8.75 &\quad Q_0^2=12\cr
\mu_\nu(1)=2.69\pm0.05&\qquad2.71\pm0.04\cr
\mu_\nu(3)=0.099\pm0.018&\qquad0.095\pm0.017\cr
\mu_\nu(5)=0.016\pm0.003&\qquad0.015\pm0.003\cr
\mu_\nu(7)=0.0042\pm0.00015&\qquad0.0039\pm0.0013\cr
\mu_\nu(9)=0.0012\pm0.0008&\qquad0.0011\pm0.0005\cr
\mu_\nu(11)=0.00036\pm0.00015&\qquad0.00032\pm0.00011.\cr
\mu_\nu(13)=0.00015\pm0.00011&\qquad0.00012\pm0.00009.\cr}$$
An important consistency check is the comparison of these with the 
nonsinglet $ep$ moments 
(after interpolating to get from odd to even moments, and multiplying by 
the average charge factor, $5/18$). We find,
for $Q_0^2=8.75$,

$$\matrix{n&(5/18) \mu(\nu N)&\mu_{NS}(ep)\cr
2&       0.110&\qquad            0.126 \pm 0.006\cr
4&       0.010&\qquad            0.0161 \pm 0.0011\cr
6&       0.0020&\qquad           0.0036 \pm 0.0005\cr
8&       0.0006&\qquad          0.0011 \pm 0.0005\cr
10&      0.00021&\qquad          0.00041 \pm 0.00017\cr
12&      0.00009&\qquad          0.00020 \pm 0.00011\cr}
$$
and, for
$Q_0^2=12$,

$$\matrix{n&   (5/18) \mu(\nu N)&      \mu_{NS}(ep)\cr
2&       0.109&\qquad            0.122 \pm 0.005\cr
4&       0.0097&\qquad           0.0150 \pm 0.0010\cr
6&       0.0019&\qquad          0.0032 \pm 0.0008\cr
8&       0.00054&\qquad          0.00090 \pm 0.0004\cr
10&      0.00020&\qquad          0.00026 \pm 0.00010\cr
12&      0.00005&\qquad          0.00009 \pm 0.00004.\cr}
$$
The discrepancies are what one would expect from corrections due to perturbative, 
experimental and interpolation effects, 
as well as the different isospin structure.

\booksubsection{3.3 Extrapolation to higher $Q^2$ for $ep$ Scattering. 
Bound on Squark and Gluino Masses}
As mentioned at the beginning of \sect~3.1 we have, 
in the calculations carried till now, only considered 
fitting averages $F_{nk}$ such that we have experimental points covering the 
whole interval 
$[\bar{x}_{nk}-\Delta\bar{x}_{nk},\bar{x}_{nk}+\Delta\bar{x}_{nk}]$ 
where the corresponding Bernstein polynomial is concentrated; 
and this, to rely only on {\sl interpolations} when evaluating the 
experimental averages. This is what limits the upper range of 
values of $Q^2$ we can use to $Q^2\leq 230\;\gev^2$. 
Now, the results of our fits are so good that we may consider extending them to 
higher values of $Q^2$. 
This will imply extrapolation to calculate the integrals, and hence larger errors for the experimental input,
as well as a  certain dependence on the parametrization 
we use to evaluate the Bernstein integrals. 
On the other hand, this will permit an important check of
 the possibility to extrapolate our results to higher $Q^2$, an important issue 
in view of eventual applications to LHC studies. 

\topinsert{
\setbox0=\vbox{\epsfxsize 8.truecm\epsfbox{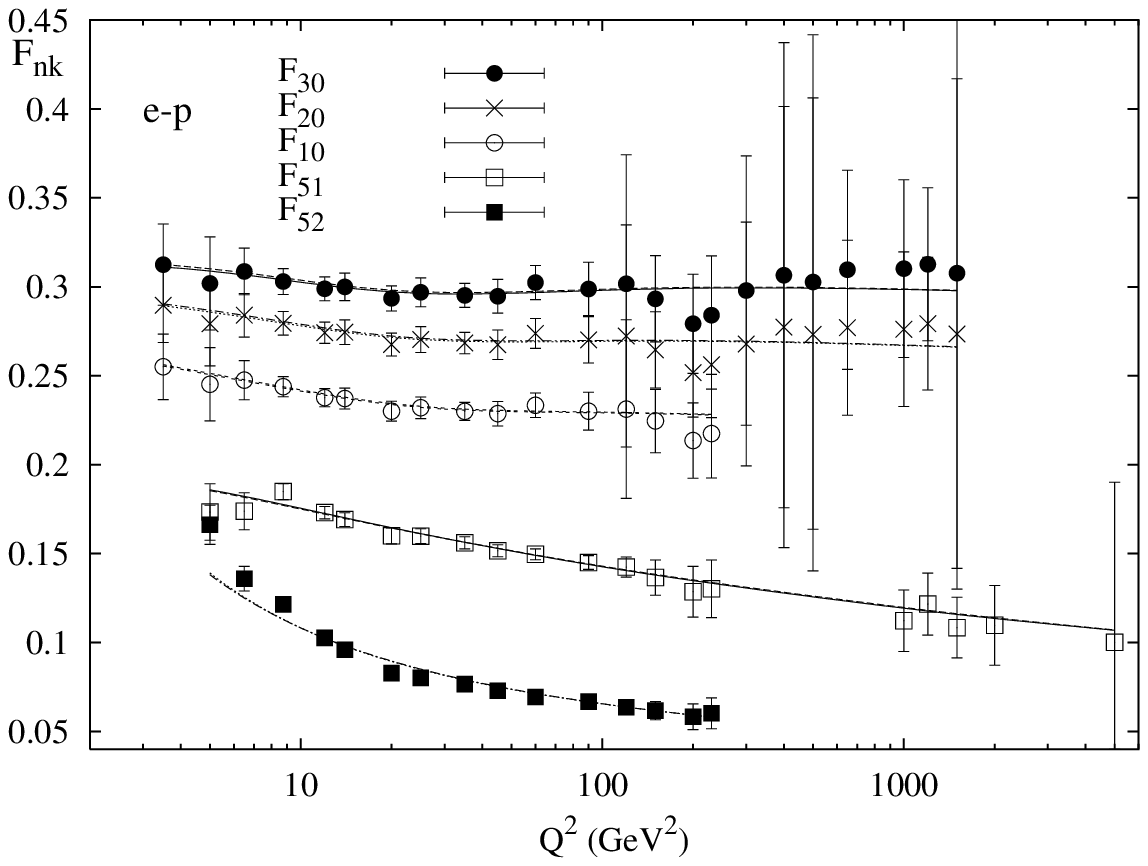}} 
\centerline{{\box0}}   
\setbox3=\vbox{\hsize 12.2truecm
\setbox0=\vbox{\epsfxsize 8.truecm\epsfbox{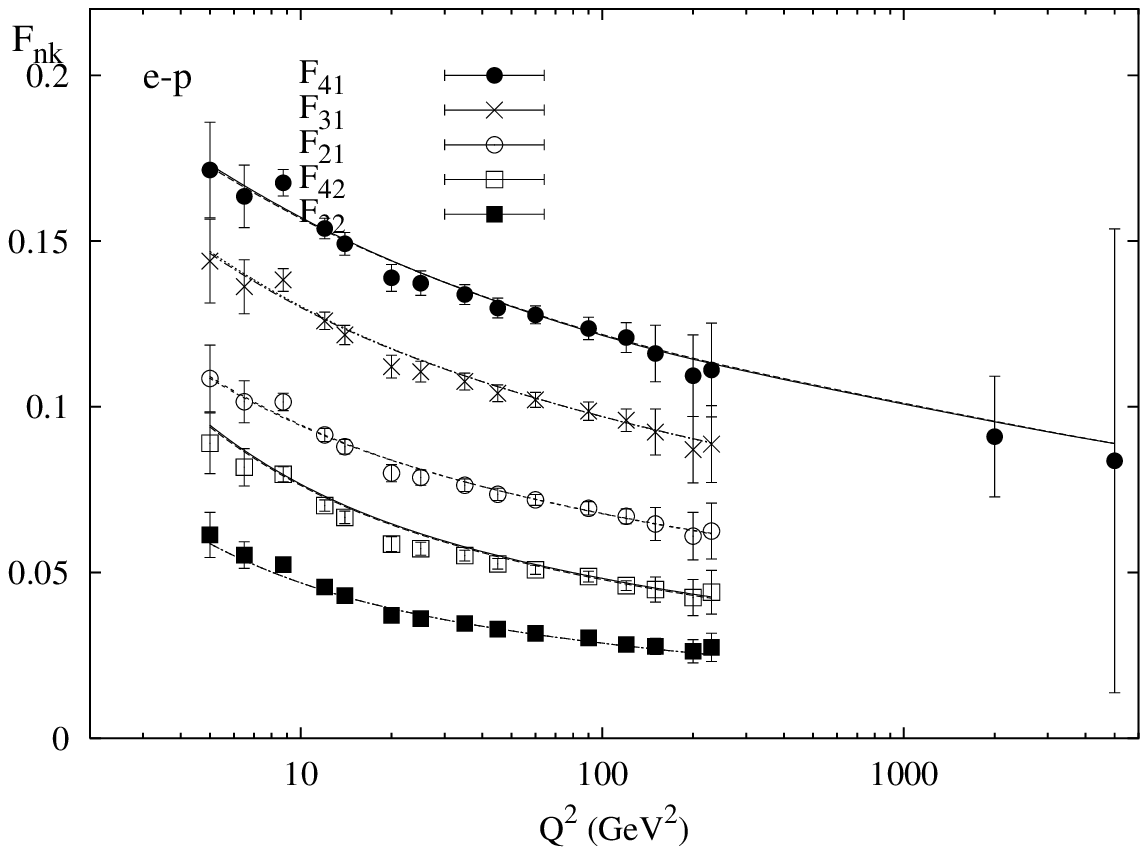}}
\setbox1=\vbox{\hsize 10.5truecm\captiontype\figurasc{Figure 3. }{Extended fit 
to $F_2$, to $Q^2$=5000 \mev.}\hb
\vskip.1cm} 
\centerline{{\box0}}
\medskip
\centerline{\box1}}
\centerline{\box 3}
\medskip
}\endinsert

To be precise, what we do is the following. 
We fit the recent HERA data on  large $Q^2$  $ep$ scattering 
for $F_2$ with an expression like the one used before, 
$$\eqalign{F_S^{\rm phen.}(x)=&(Ax^{-0.44}+C)(1-x)^\nu,\cr
F_{NS}^{\rm phen.}(x)=&B x^{0.5}(1-x)^\mu.\cr}$$
 We now admit a number of new averages, 
where the corresponding Bernstein polynomials are peaked around $x\sim0$. 
Specifically, we now admit averages provided only that at least {\sl six} 
experimental points lie inside the interval 
$[\bar{x}_{nk}-\Delta\bar{x}_{nk},\bar{x}_{nk}+\Delta\bar{x}_{nk}]$, 
(but we do not 
require that they extend to the edges) so that the fit be meaningful. 
In this way, we are able to
  include experimental averages up to $Q^2=650\;\gev^2$ for 
$F^{\rm exp.}_{20},\;F^{\rm exp.}_{30}$ 
and up to $Q^2=5\,000\;\gev^2$ for 
$F^{\rm exp.}_{41},\;F^{\rm exp.}_{51}$;  
other averages would get so large errors to make them meaningless. 
We then follow two strategies: i) We {\sl include} the new points in the fit, 
which tests the stability of our results. 
If we do so we obtain the new results,
$$\eqalign{\Lambdav(n_f=4,\,\hbox{3 loop})=&269\pm14\;\mev\;(\hbox{stat.});\cr 
 \alpha_s^{(\rm 3\, loop)}(M_Z)=&0.1163\pm0.0010\;(\hbox{stat.})\cr} 
\equn{(3.7)}$$
with a \chidof\ of $\chidof=93.5/(174-18)=0.60$
slightly better than with the restricted fit.\fnote{We can even interpret the slightly 
smaller value of $\lambdav$ obtained here as due to the 
greater weight that have now large $Q^2$ data, so diminishing the 
contamination with higher twist effects.} 
The agreement of (3.7) with (3.5) proves the noteworthy stability of our results.

ii) Alternatively, we can {\sl extrapolate} our results, with the central values as given 
in \equn~(3.5),  plot the corresponding values for (the averages of) 
$F_2$, and compare them both with data and with the curves
 obtained from the overall fit obtained following 
the procedure outlined in (i) above. 

The theoretical curves obtained with both methods, 
together with the corresponding experimental values,  
are shown in \fig~3. The agreement of the curves found fitting also
 the new points, or simply extrapolating our previous results 
is so good you can hardly tell them apart in the figure. 
This shows very clearly the {\sl predictive} power of our evaluation: 
from the input region, between 3.5 and $230\;\gev^2$, to the $5\,000\;\gev^2$ 
of \fig~3.

As an application of this, we improve on the bounds on squark and gluino masses 
obtained in ref.~2. Repeating the analysis carried there, we may conclude that the 
fact that the \chidof\ actually 
{\sl decrease} (slightly) when including high $Q^2$ data implies that no new 
particles are excited in this range. 
This provides the bound
$$m_{\tilde{q}},\,m_{\tilde{g}}\geq\,60\;\gev$$
 for the masses of squarks or gluinos.

\booksection{4. Comparison with other calculations}
We next show a table comparing our results to previous determinations
 for $\alpha_s(M_Z^2)$. These are taken from the review by Bethke\ref{14}, 
except the $Z\to{\rm jets}$ one which is from ref.~15 and 
the one referred to as ``our old result" which is of course that of ref.~1.
We include only the processes where the theoretical calculation
 has been pushed to the NNLO level.

\setbox0=\vbox{
\medskip
\setbox1=\vbox{\offinterlineskip\hrule
\halign{
&\vrule#&\strut\hfil#\hfil&\quad\vrule\quad#&
\strut\quad#\quad&\quad\vrule#&\strut\quad#\cr
 height2mm&\omit&&\omit&&\omit&\cr 
& \kern.5em Process&&${\textstyle\hbox{Average}\;
 Q^2}\atop{\textstyle \hbox{or}\; Q^2\;\hbox{range}\;[\gev]^2}$&
& $\alpha_s(M_Z^2)$\kern.3em& \cr
 height1mm&\omit&&\omit&&\omit&\cr
\noalign{\hrule} 
height1mm&\omit&&\omit&&\omit&\cr
&\phantom{\Big|}  DIS; $\nu$, Bj&&1.58&&$0.121^{+0.005}_{-0.009}$\phantom{l}& \cr
\noalign{\hrule}
&\phantom{\Big|}  DIS; $\nu$, GLS&&3&&$0.112\pm0.010$\phantom{l}& \cr
\noalign{\hrule}
&\phantom{\Big|}  $\tau$ decays&&$(1.777)^2$&&$0.1181\pm0.0031$&\cr
\noalign{\hrule}
&\phantom{\Big|}  $e^+e^-\to{\rm hadrons}$&&$100 - 1600$&&$0.128\pm0.025$&\cr
\noalign{\hrule}
&\phantom{\Big|}  $Z\to{\rm hadrons}$&&$(91.2)^2$&&$0.121^{+0.005}_{-0.004}$&\cr  
\noalign{\hrule}
&\phantom{\Big|}  $Z\to{\rm jets}$&&$(91.2)^2$&&0.117&\cr
\noalign{\hrule}
&\phantom{\Big|}  Our result($\nu\,N;\;xF_3$)&&$8 - 120$&&$0.1153\pm0.0041$\phantom{l}& \cr
\noalign{\hrule}
&\phantom{\Big|}  Our {\sl old} result ($ep$)&&${2.5} - {230}$&&$0.1172\pm0.0024$\phantom{l}&\cr
\noalign{\hrule}
&\phantom{\Big|}  Our result now ($ep$, I)&&${3.5} - {230}$&&$0.1166\pm0.0013$\phantom{l}&\cr
\noalign{\hrule}
&\phantom{\Big|}  Our result now ($ep$, II)&&${3.5} - {5000}$&&$0.1163\pm0.0014$\phantom{l}&\cr
 height1mm&\omit&&\omit&&\omit&\cr
\noalign{\hrule}}
\vskip.05cm}
\centerline{\box1}
\smallskip
\centerline{\petit Table 5}
\centerrule{6cm}
\medskip}
\box0
\noindent Here DIS means deep inelastic scattering, 
Bj stands for the Bjorken, and GLS for the Gross--Llewellyn Smith 
sum rules. 
We have presented our two results here, the one for the smaller $Q^2$ range (I, that 
we consider the more reliable) and the extended range one (II). 
The value given for $Z\to \hbox{hadrons}$ assumes the Higgs mass 
constrained by $115\leq M_H\leq 300\;\gev$, with the central value for $M_H=115\;\gev$. 
  All determination are compatible with 
one another, within errors. 

We finish this paper with brief comments on other recent work 
on our subject. First, we have the evaluations of Kataev et al.\ref{16} 
who consider $\nu N$ scattering to NNLO. In 
their work, it is necessary to interpolate to obtain even moments and also to 
assume a Jacobi polynomial expansion for the structure function. 
They find the value $\lambdav(n_f=4,\;3\;\hbox{loop})=332\pm36$, 
where some systematic errors are not taken into account. 
The result is compatible with ours, within errors. 
A second paper is that of van Neerven and Vogt\ref{17} 
who, however, are only interested in the consistency of the data with NNLO 
perturbation theory but, 
as far as we can see, no effort is made to extract the value of $\alpha_s$.

\booksection{Appendix}
The constraints imposed on the moments of a positive function are 
based on the following result:
\medskip
\noindent {\bf Theorem (Lukacs)}.
A polynomial positive in the interval  $[0,+1]$ may be written as
$$P_{2n}(x)=
\left(\sum_{k=0}^{n}a_kx^k\right)^2+
x(1-x)\left(\sum_{k=0}^{n-1}b_kx^k\right)^2$$
(even degree)
$$P_{2n-1}(x)=
x\left(\sum_{k=0}^{n-1}a_kx^k\right)^2+
(1-x)\left(\sum_{k=0}^{n-1}b_kx^k\right)^2$$
(odd degree). 
See e.g. Ahiezer and Krein, ref.~8,  p.~35.

For polynomials in $x^2$, we write $P(x^2)$, 
$P$ as before; likewise, $xP(x^2)$ gives 
polynomials in odd powers of $x$.
\smallskip
\noindent{\bf Positivity}

Consider the moments
$$M_i=\int_0^1\dd t\,t^i\Psi(t),\quad i=0,1,\dots,N,\equn{(A1)}$$
where the function $\Psi$ is positive ({\sl Hausdorff moment  problem}). 
We are going to describe necessary and sufficient conditions 
for the $M_i$ to verify (A1). The conditions 
depend on whether $N$ is even or odd, and are derived from the
representation of polynomials positive in the interval $[0,1]$.

{\bf Theorem}. i) Let $N=2m$ be even. Then, if $P_N$ is positive 
we write it  as
$$P_N(t)=\left(\sum_{i=0}^{m}a_it^i\right)^2+
t(1-t)\left(\sum_{i=0}^{m-1}b_it^i\right)^2.\equn{(A2a)}$$

ii) If $N=2m-1$ is odd,
$$P_N(t)=t\left(\sum_{i=0}^{m-1}a_it^i\right)^2+
(1-t)\left(\sum_{i=0}^{m-1}b_it^i\right)^2.\equn{(A2b)}$$
On integrating the $P_N$ with $\Psi$ we find the following conditions:

i) For $N$ even:
$$\sum_{i,j=0}^ma_ia_jM_{i+j}+
\sum_{i,j=0}^{m-1}b_ib_j[M_{i+j+1}-M_{i+j+2}]\geq0;\equn{(A3a)}$$
for $N=$odd,
$$\sum_{i,j=0}^{m-1}a_ia_jM_{i+j+1}+
\sum_{i,j=0}^{m-1}b_ib_j[M_{i+j}-M_{i+j+1}]\geq0.\equn{(A3b)}$$

The problem is solved completely by using the
following result:

\noindent{\bf Theorem}. The quadratic form
$$Q[\xi]=\sum_{i,j=0}^n\lambda_{i+j}\xi^i\xi^j$$
is positive if, and only if, all determinants
$$\Delta_{k}[\lambda]=\det\pmatrix{\lambda_0&\lambda_{1}&\dots&\lambda_{k}\cr
\lambda_{1}&\lambda_{2}&\dots&\lambda_{k+1}\cr
\dots&\dots&\dots&\dots\cr
\lambda_{k}&\lambda_{k+1}&\dots&\lambda_{2k}\cr}
\equn{(A4})$$
are positive for $2k\leq n$.
\smallskip
\noindent{\bf Connection with DIS}

For electroproduction, and $F=F_{NS},\,F_S,\,F_G$ you know
$$\mu_n=\int_0^1\dd x\,x^{n-2}F(x),\quad n=2,4,6,\dots.$$
Set then 
$$t=x^2,\quad n-2=2i,\quad \Psi(t)=\tfrac{1}{2}t^{-1/2}F(t^{1/2});$$
then you have the $M_i$ given by
$$M_i=\int_0^1\dd t \,t^i\Psi(t), \quad i=0,1,2\dots$$
For neutrino scattering,
$$\mu_n=\int_0^1dx\,x^{n-1}F_3(x),\quad n=1,3,5,\dots.$$
With $n-1=2i$ now, the rest as before, we get moments
$$M_i=\int_0^1dx \,t^i\Psi(t), \quad i=0,1,2\dots$$

\vfill\eject
\booksection{Acknowledgments}
The authors are indebted to A. Vermaseren and G.~Parente  for
 several illuminating discussions. One of us wishes to thank the 
Schr\"odinger Institute, where  part of this work was made.

 The financial support of CICYT, MCYT (FPA2000-1558) and Junta de Andaluc\'{\i}a 
(FQM-101) (Spain), 
and of the European Union (HPRN-CT-2000-00149) is  gratefully acknowledged. 

\booksection{References}
\item{1.- }{\ajnyp{J.~Santiago and F. J. Yndur\'ain}{Nucl. Phys.}{B563}{1999}{45}}
\item{2.- }{\ajnyp{J.~Santiago and F. J. Yndur\'ain}{Nucl. Phys. B (Proc. Suppl.)}{86}{2000}{69}}
\item{3.- }{\ajnyp{W. L. van Neerven and E. B. Zijlstra}
{Phys. Lett.}{B272}{1991}{127 and 476}; 
{\sl ibid} { B273} {(1991)} {476}; {\sl Nucl. Phys.} {\bf B383} (1992) 525.}
\item{4.- }{\ajnyp{S. A. Larin et al.}{Nucl. Phys.}{B427}{1994}{41} and 
{\bf B492} (1997), 338.}
\item{5.- }{\ajnyp{F. J. Yndur\'ain}{Phys. Lett.}{74B}{1978}{68}.}
\item{6.- }{{\sc  A. Rety and J. A. M. Vermaseren}, hep-ph/0007294.} 
\item{7.- }{{\sc J. A. Shohat and J. D. Tamarkin}, ``The Problem of Moments'', 
Amer. Math. Soc., 1943.}
\item{8.- }{{\sc N. I. Akhieser and M. Krein}, ``Some Questions in the Theory of Moments'', 
Amer. Math. Soc., 1962.}
\item{9.- }{For $ep$ data: \ajnyp{L. W. Whitlow et al.}{Phys. Lett.}{B282}{1992}{475};
\ajnyp{A. Benvenuti et al.}{Phys. Lett.}{B223}{1989}{485}; 
\ajnyp{M. R. Adams et al.}{Phys. Rev.}{D54}{1996}{3006}; 
\ajnyp{M. Derrick et al.}{Z. Phys.}{C72}{1996}{399};  
\ajnyp{S. Aid et al.}{Nucl. Phys.}{B470}{1996}{3}; 
\ajnyp{Adloff et al.}{Eur. Phys. J.}{C13}{2000}{609}.} 
\item{10.- }{\ajnyp{W. G. Seligman et al.}{Phys. Rev. Lett.}{79}{1997}{1213}.}
\item{11.- }{\ajnyp{T. van Ritbergen, J. A. M. Vermaseren and
 S. A. Larin}{Phys. Lett.}{B400}{1997}{379}.}
\item{12.- }{\ajnyp{G. K. Chetyrkin, B. A. Kniehl and
 M. Steinhauser}{Phys. Rev. Lett.}{79}{1997}{2184}.}
\item{13.- }{\ajnyp{A. Pineda and F. J. Yndur\'ain}{Phys. Rev.}{D58}{1998}{094022} 
and ibid., {\bf D61}, 077505 (2000).}
\item{14.- }{\ajnyp{S. Bethke}{J. Phys.}{G26}{2000}{R27}.}
\item{15.- }{\ajnyp{P. Abreu et al.}{Eur. Phys. J.}{C14}{2000}{557}.}
\item{16.- }{\/ {\sc A. L. Kataev, G. parente and A. V. Sidorov}, hep-ph/0012014.}
\item{17.- }{\ajnyp{W. L. van Neerven and
A. Vogt}{Nucl. Phys.}{B588}{2000}{345}; {\sl Phys. Lett.} {\bf B490}
(2000) 111.}
\item{}{}

\bookendchapter


\bye